\documentclass[preprint,journal]{vgtc}            

\onlineid{1792}

\vgtccategory{Representations \& Interaction}

\title{Juicy Interactive Visualization: Evaluating How Excessive Feedback Design Shapes Visualization Engagement}

\author{%
    \authororcid{Shano Liang}{0000-0003-1976-9680},
    \authororcid{Max Chen}{0000-0002-2738-6017}, and
    \authororcid{Lane Harrison}{0000-0003-3029-2799}
}

\authorfooter{
  \item Shano Liang is with Worcester Polytechnic Institute. 
  E-mail: sliang1@wpi.edu.
  \item Max Chen is with Worcester Polytechnic Institute. 
  E-mail: mjchen@wpi.edu.
  \item Lane Harrison is with Worcester Polytechnic Institute. 
  E-mail: ltharrison@wpi.edu.
}

\abstract{%
    Visualization research has long examined embellishment and animation, while the design of rich interaction-contingent feedback remains under-articulated as a systematic design space. 
    We introduce \textit{JuicyVIS}, a theory-informed operationalization of juicy feedback for interactive visualization that translates ideas from game studies into a visualization-centered framework. JuicyVIS introduces three dimensions, interaction type, feedback timing, and feedback intensity, 
    which we explore through 26 controlled prototypes, grounded in established visualization interaction categories and prior work on juicy feedback in games. We evaluate the juicy design space through three exploratory mixed-methods online studies. 
    Results show that juicy feedback was consistently associated with higher engagement and aesthetic experience, while the clearest gains came from post-interaction feedback, not, for example, by simply increasing intensity. 
    We discuss how juicy feedback can be thought of as a design material in visualizations, shaping how visualization interactions feel, how action consequences are perceived, and how users remain engaged with data representations.
}


\teaser{
  \centering
  \includegraphics[width=0.92\linewidth]{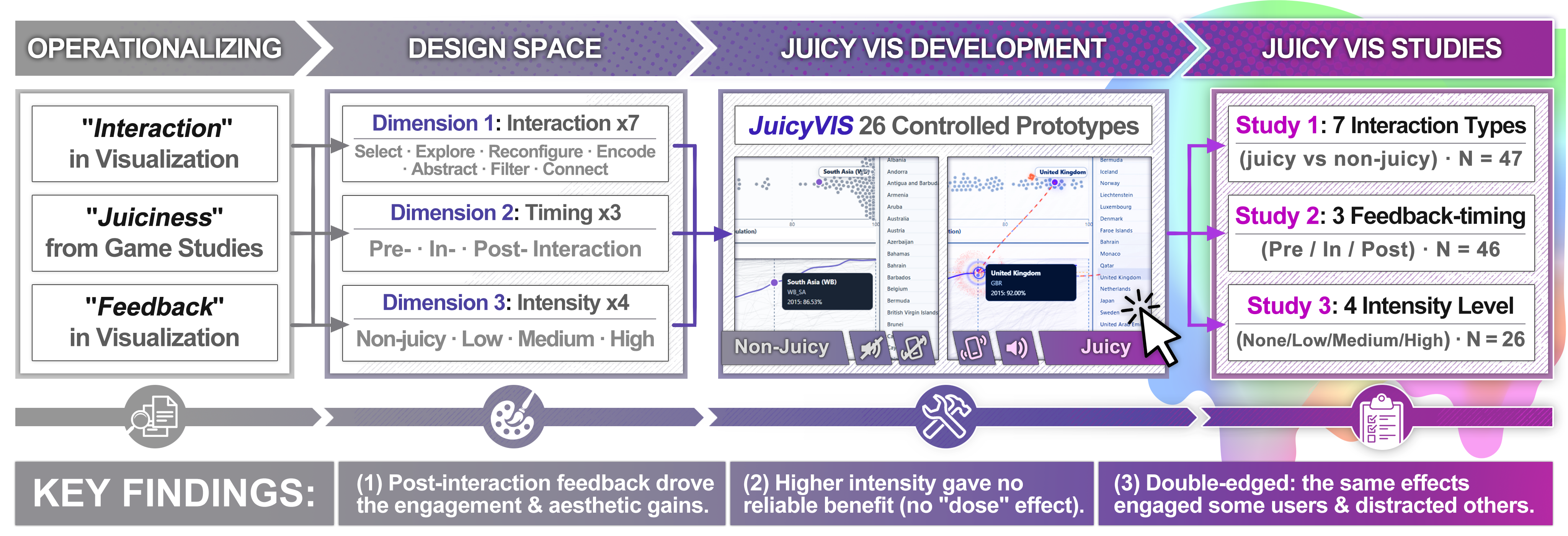}
  \caption{%
  	Overview of our work. 
    We operationalize game-studies juiciness as a visualization design space. 
    Grounding a palette of juicy feedback techniques (\cref{tab:juicepalette}), we instantiate three dimensions (interaction type, feedback timing, and intensity) across 26 controlled prototypes and evaluate them in three exploratory online studies.%
  }
  \label{fig:teaser}
}

\graphicspath{{figs/}{figures/}{pictures/}{images/}{./}} 

\usepackage{tabu}                      
\usepackage{booktabs}                  
\usepackage{threeparttable}
\usepackage{lipsum}                    
\usepackage{mwe}                       
\usepackage{ccicons}                   
\usepackage{colortbl}
\usepackage{array}
\usepackage{arydshln} 
\usepackage{mathptmx}                  

\usepackage{multirow}

\usepackage{tabularx}

\definecolor{JuiceHeader}{rgb}{0.91,0.91,0.91}

\definecolor{JuiceAnimDark}{rgb}{0.357,0.165,0.659}
\definecolor{JuiceAnimLight}{rgb}{0.941,0.910,0.980}

\definecolor{JuiceParticleDark}{rgb}{0.122,0.361,0.749}
\definecolor{JuiceParticleLight}{rgb}{0.910,0.941,1.000}

\definecolor{JuiceAudioDark}{rgb}{0.157,0.478,0.227}
\definecolor{JuiceAudioLight}{rgb}{0.906,0.953,0.910}

\definecolor{JuiceShakeDark}{rgb}{0.604,0.306,0.000}
\definecolor{JuiceShakeLight}{rgb}{1.000,0.941,0.871}

\definecolor{JuicePersistDark}{rgb}{0.722,0.196,0.141}
\definecolor{JuicePersistLight}{rgb}{0.984,0.906,0.890}

\newcommand{\JuiceTag}[2]{%
  \begingroup
  \setlength{\fboxsep}{1.15pt}%
  \colorbox{#1}{%
    \strut
    \color{white}%
    \sffamily\bfseries
    \fontsize{7.3pt}{7.8pt}\selectfont
    #2%
  }%
  \endgroup
}

\newcommand{\jA}[1]{\JuiceTag{JuiceAnimDark}{A\_#1}}
\newcommand{\jPa}[1]{\JuiceTag{JuiceParticleDark}{Pa\_#1}}
\newcommand{\jAu}[1]{\JuiceTag{JuiceAudioDark}{Au\_#1}}
\newcommand{\jS}[1]{\JuiceTag{JuiceShakeDark}{S\_#1}}
\newcommand{\jP}[1]{\JuiceTag{JuicePersistDark}{P\_#1}}

\newcommand{\JuiceTerm}[4]{%
  \begingroup
  \setlength{\fboxsep}{0.60pt}%
  \mbox{%
    \colorbox{#1}{\strut #3}%
    \nobreakspace
    \JuiceTag{#2}{#4}%
  }%
  \endgroup
}

\newcommand{\jATerm}[2]{%
  \JuiceTerm{JuiceAnimLight}{JuiceAnimDark}{#1}{A\_#2}%
}
\newcommand{\jPaTerm}[2]{%
  \JuiceTerm{JuiceParticleLight}{JuiceParticleDark}{#1}{Pa\_#2}%
}
\newcommand{\jAuTerm}[2]{%
  \JuiceTerm{JuiceAudioLight}{JuiceAudioDark}{#1}{Au\_#2}%
}
\newcommand{\jSTerm}[2]{%
  \JuiceTerm{JuiceShakeLight}{JuiceShakeDark}{#1}{S\_#2}%
}
\newcommand{\jPTerm}[2]{%
  \JuiceTerm{JuicePersistLight}{JuicePersistDark}{#1}{P\_#2}%
}

\newcommand{\JuiceGroup}[3]{%
  \noalign{\vskip 2pt}%
  \rowcolor{#1}%
  \multicolumn{1}{%
    @{}%
    >{\columncolor{#1}\color{white}\centering\arraybackslash}%
    m{1.85cm}%
  }{%
    \fontsize{7.3pt}{7.9pt}\selectfont
    \sffamily\bfseries
    #2%
  }%
  &
  \multicolumn{2}{%
    >{\columncolor{#1}\color{white}\raggedright\arraybackslash}%
    m{\dimexpr\columnwidth-1.85cm-4\tabcolsep\relax}%
    @{}%
  }{%
    \fontsize{6.9pt}{7.5pt}\selectfont
    \sffamily
    #3%
  }%
  \\%
  \noalign{\vskip 2pt}%
}

\newcommand{\JuicePaletteRow}[4]{%
  \rowcolor{#1}%
  \textbf{#2}%
  &
  #3%
  &
  #4%
  \\%
  \arrayrulecolor{white}\hline
}

\usepackage{tabularx}
\usepackage{soul}

\providecommand{\JuiceInlineTag}[2]{}
\renewcommand{\JuiceInlineTag}[2]{%
  \begingroup
  \setlength{\fboxsep}{0.75pt}%
  \raisebox{0.05ex}{%
    \colorbox{#1}{%
      \color{white}%
      \sffamily\bfseries
      \fontsize{5.8pt}{6.1pt}\selectfont
      \strut #2%
    }%
  }%
  \endgroup
}

\providecommand{\JuiceTerm}[4]{}
\renewcommand{\JuiceTerm}[4]{%
  \begingroup
  \sethlcolor{#1}%
  \hl{#3}%
  \endgroup
  \nobreakspace
  \JuiceInlineTag{#2}{#4}%
}

\providecommand{\jATerm}[2]{}
\renewcommand{\jATerm}[2]{%
  \JuiceTerm{JuiceAnimLight}{JuiceAnimDark}{#1}{A\_#2}%
}

\providecommand{\jPaTerm}[2]{}
\renewcommand{\jPaTerm}[2]{%
  \JuiceTerm{JuiceParticleLight}{JuiceParticleDark}{#1}{Pa\_#2}%
}

\providecommand{\jAuTerm}[2]{}
\renewcommand{\jAuTerm}[2]{%
  \JuiceTerm{JuiceAudioLight}{JuiceAudioDark}{#1}{Au\_#2}%
}

\providecommand{\jSTerm}[2]{}
\renewcommand{\jSTerm}[2]{%
  \JuiceTerm{JuiceShakeLight}{JuiceShakeDark}{#1}{S\_#2}%
}

\providecommand{\jPTerm}[2]{}
\renewcommand{\jPTerm}[2]{%
  \JuiceTerm{JuicePersistLight}{JuicePersistDark}{#1}{P\_#2}%
}

\providecommand{\jAInline}[1]{}
\renewcommand{\jAInline}[1]{%
  \JuiceInlineTag{JuiceAnimDark}{A\_#1}%
}

\providecommand{\StudyOnePrototypeLinks}[1]{}
\renewcommand{\StudyOnePrototypeLinks}[1]{%
  \begin{tabular}[c]{@{}l@{}}
    \href{https://juicy-vis-10aac.web.app/?chart=#1\&juicy=1}
      {\textbf{[Juicy\(\nearrow\)]}}
    \\[1.5pt]
    \href{https://juicy-vis-10aac.web.app/?chart=#1\&juicy=0}
      {\textbf{[Non-juicy\(\nearrow\)]}}
  \end{tabular}%
}

\providecommand{\StudyTwoPrototypeLink}[2]{}
\renewcommand{\StudyTwoPrototypeLink}[2]{%
  \href{https://juicy-vis-10aac.web.app/?chart=8\&juicy=#1}
    {\textbf{[#2\(\nearrow\)]}}%
}

\usepackage{mathptmx}                  

\begin{document}


\firstsection{Introduction}

\maketitle

``\textit{Juice},'' or ``\textit{juiciness},'' a concept widely adopted in game and interactive media studies, refers to the quality of excessive amounts of feedback in relation to user input \cite{kao2024does}, while design and feedback that exhibit this quality are considered ``\textit{\textbf{juicy}}.'' 
Juiciness changes how actions feel rather than what an interface can do, amplifying the perceived responsiveness, salience, and ``liveness'' of its reactions to input. 
It has been used to explain why seemingly small micro-feedback can motivate continued interaction
\cite{hicks2018good,hicks2019juicy,kao2020effects,kao2024does}.

Visualization research, meanwhile, has long examined forms of ``extra'' presentation beyond minimalism, including static embellishment \cite{bateman2010useful} and animation techniques \cite{amini2018hooked,heer2007animated,robertson2008effectiveness}. 
Many works treat such effects primarily as presentation-level styling; far fewer studies frame them as interaction-contingent feedback tightly coupled to moment-to-moment user input, that is, what a visualization \textit{richly} ``says back'' as users act on it. 
In parallel, visualization interaction researchers have argued that feedback is central to effective interaction \cite{elmqvist2011fluid},
yet few studies have systematically mapped feedback-centric design choices onto established interaction categories and then varied feedback properties as primary experimental factors.
As a result, the quality of interaction-time visualization feedback remains under-articulated as a systematic design space: many systems are functionally correct yet respond so minimally that interaction feels experientially ``thin.''

Recent work by Campos has begun to bring juiciness into the visualization field, through a content-analysis survey of effect categories in existing infographics \cite{campos2025assessing}, a designer ethnography \cite{campos2025juiciness}, and a comparative juicy-versus-dry study \cite{campos2025juicy}. 
This work establishes that juicy techniques already circulate in visualization practice, while operating largely at a descriptive, representation-level register. What remains open is a theory-grounded design space that ties juicy feedback to specific interaction events and treats it as a primary design factor.

Games have long treated interaction quality as a core design material. 
We turn to game studies for a vocabulary suited to this gap, taking juicy feedback design as a lens for articulating
the experiential ``\textit{thickness}'' of visualization feedback. 
We therefore ask: how can juiciness be translated from game studies into a visualization-centered design space for interaction feedback, and how do its dimensions relate to users' experiences and task-related outcomes in interactive visualizations?

Building on this prior groundwork, we present \textit{JuicyVIS} (\cref{fig:teaser}), a theory-informed operationalization of juicy interaction feedback for data visualization. 
We map juiciness onto established visualization interaction types \cite{yi2007toward} and instantiate three design dimensions: \textit{\textbf{interaction type}}, \textit{\textbf{feedback (timing) type}}, and \textit{\textbf{feedback intensity}}, across \textbf{26} controlled prototypes.
We then conduct three exploratory mixed-methods online studies that pair comparative quantitative measures with qualitative accounts, examining how these dimensions relate to engagement, aesthetic experience, preference, and task-related outcomes.
Across these studies, juicy feedback was consistently associated with higher engagement and aesthetic experience, with the clearest gains from post-interaction feedback and no simple ``\textit{more is better}'' effect for intensity. 
Our aim is to make this under-articulated design space practically and empirically inspectable, and to surface promising patterns, trade-offs, and boundary conditions for future visualization design.

\section{Background}

Both \textit{\textbf{juiciness}} in game studies and \textit{\textbf{interaction}} in visualization have been discussed in prior literature and widely used as umbrella terms, yet their concepts are usually ambiguous and inconsistently defined across studies. Therefore, 
we first review literature and articulate how these concepts are contextualized and operationalized in our study to clarify the scope of using and constructing ``\textit{juiciness}'' and ``\textit{interaction}.''

\subsection{Operationalizing \textit{Interaction} in Visualization}

Because juiciness concerns input-contingent feedback, studying juicy feedback in visualization first requires specifying which user actions constitute interaction events and thus can be ``juicified.'' 
For this work's purposes, we seek an interaction taxonomy that is sufficiently actionable to serve as the framework for constructing controlled juicy design.

Two frameworks are mostly relevant: Yi et al. \cite{yi2007toward} propose a grounded, intent-oriented taxonomy of seven interaction archetypes.
In parallel, Dimara and Perin \cite{dimara2019interaction} synthesize an intentionally inclusive view of interaction for data visualization, framing interaction as a broad space of actions across the visualization pipeline.
Other adjacent works of visualization research conceptualize interaction at different analytic resolutions. 
Steed \cite{steed2025interactive} approaches interactive visualization from a pipeline and workflow perspective, emphasizing that user intervention may occur at multiple stages. 
Amar et al. \cite{amar2005low}, by contrast, focus on the analytic work users seek to accomplish, proposing ten low-level analytic tasks as a common vocabulary. 
Brehmer and Munzner \cite{brehmer2013multi} similarly introduce
a multi-level typology that distinguishes why a task is undertaken and how it is carried out.
These perspectives are valuable for situating interaction within broader analytic activity. 
Yet, because they are not suitable for serving as an input-event-level operational backbone for operationalizing interaction feedback, we treat them here as contextual lenses. 

\begin{table}[tb]
\renewcommand{\arraystretch}{1.1}
  \caption{%
  	Comparison of Yi et al.'s \cite{yi2007toward} and Dimara and Perin's \cite{dimara2019interaction} visualization interaction taxonomies. The pairings shown here represent interpretive, closest correspondences rather than exact matches.
  }
  \label{tab:vis_interaction}
  \scriptsize%
	\newlength{\digitwidth}%
  \settowidth{\digitwidth}{0}%
  \centering%
  \begin{tabular}{%
          @{\hspace{4pt}}>{\raggedright\arraybackslash}m{4.2cm}%
          @{\hspace{4pt}}>{\columncolor[gray]{0.94}[3pt][5pt]\raggedright\arraybackslash}m{4.1cm}%
  	}
  	\toprule
  	\centering \textbf{Yi et al.'s taxonomy} \cite{yi2007toward} & \textbf{Dimara and Perin's taxonomy} \cite{dimara2019interaction} \\
  	\midrule
    
    \multicolumn{1}{c}{\textbf{[No direct counterpart]}} & 
    \textbf{(1.1) Input Data Actions}: ``operate on raw data values'' \\
        \hdashline
        
    \textbf{(6) Filter}: ``show me something conditionally'' (e.g., control to show/hide subsets) & 
    \textbf{(1.2) Processing Data Actions}: ``apply functions/transformations to the input/raw data'' (e.g., compile) \\
    
    \hdashline
    \begin{tabular}[t]{@{}>{\raggedright\arraybackslash}p{4.1cm}@{}}
    \textbf{(3) Reconfigure}: ``show me a different arrangement'' (e.g., reorder) \\
        \cdashline{1-1}
    \textbf{(4) Encode}: ``show me a different representation'' (e.g., change visual encodings such as color or shape)
    \end{tabular}
    &
    \textbf{(2.1) Mapping Data Actions}: ``assign data entities to perceptual marks and variables, and layout/rearrange the data'' \\
    
            \hdashline
        \begin{tabular}[t]{@{}>{\raggedright\arraybackslash}p{4.1cm}@{}}
        \textbf{(1) Select}: ``mark something as interesting'' (e.g., mark a data item) \\
            \cdashline{1-1}
        \textbf{(2) Explore}: ``show me something else'' (e.g., move to another subset) \\
            \cdashline{1-1}
        \textbf{(5) Abstract / Elaborate}: ``show me more or less detail'' (e.g., drill down) \\
            \cdashline{1-1}
        \textbf{(6) Filter}: ``show me something conditionally''\\    
            \cdashline{1-1}
        \textbf{(7) Connect}: ``show me related items'' (e.g., highlight linked items) \\      
        \end{tabular}
        &
        
    \textbf{(2.2) Presentation Data Actions}: ``operate on the specific presentation of the data'' (e.g., mark something as interesting; navigate; stylize; highlight; decorate) \\
    
        \hdashline
    \multicolumn{1}{c}{\textbf{[No direct counterpart]}} & 
    \textbf{(3.1) Meta Actions}: ``operate on the person's own actions'' 
    \\
    
        \hdashline
    \multicolumn{1}{c}{\textbf{[No direct counterpart]}} & 
    \textbf{(3.2) Social Actions}: ``connect the person to other persons'' \\
    
        \hdashline
    \multicolumn{1}{c}{\textbf{[No direct counterpart]}} & 
    \textbf{(3.3) Interface Actions}: ``operate on other data interface components'' 
    \\
  	\bottomrule
  \end{tabular}%
  \vspace{-14pt}
\end{table}

As \cref{tab:vis_interaction} shows, Yi et al.'s seven intent categories \cite{yi2007toward} fall largely within the portion of Dimara and Perin's framework \cite{dimara2019interaction} concerned with perceptualization: actions on how data are represented, arranged, and presented, with \textit{Filter} also bordering processing-oriented actions. 
This pattern follows from the different organizing logics of the two frameworks. 
Yi et al. classify interaction by user intent, offering seven categories for characterizing what users seek to accomplish through interaction. 
Dimara and Perin, by contrast, broaden the scope of visualization interaction and organize it by where actions occur semantically in the visualization pipeline: on raw data (i.e., the ``\textbf{\textit{Data Actions}}'', including \textit{Input Data Actions} and \textit{Processing Data Actions}), on its perceptualization (including \textit{Mapping Data Actions} and \textit{Presentation Data Actions}), or on non-data entities (i.e., ``\textbf{\textit{Non-Data Actions}}'', including \textit{Meta Actions}, \textit{Social Actions}, and \textit{Interface Actions}). 
This comparison clarifies the scope of juicy interaction in the present work. 
We exclude the non-data actions because they concern meta-level activity, social coordination, and general interface management rather than direct interaction with representations. 
More importantly, Dimara and Perin distinguish between actions on raw data itself and actions on its perceptualization. 
For our purposes, we focus on the latter: input-contingent feedback tied to representation-level interactions experienced by users within an already instantiated visualization. 
Accordingly, Yi et al.'s taxonomy provides a more suitable operational framework for constructing controlled stimuli. 

\subsection{Operationalizing \textit{Juiciness} from Game Studies}

Similar to the \textit{interaction} in visualization studies, juiciness is also widely invoked but inconsistently defined across game studies. 
In an early practitioner articulation, Gabler et al. describe juicy game elements as ones that ``will bounce and wiggle and squirt and make a little noise when you touch it'' \cite{kucic2005howto}. 
This description defines juiciness from a person-internal experiential intuition, while subsequent work retains this general intuition and expands the definition (but diverges in emphasis). 
Longanecker and Brown \cite{longanecker2005howto}, for example, frame juiciness from the perspective of experiential results, claiming the juicy feedback is an interaction quality to elicit visceral and  highly satisfying responses.
Hicks et al. describe juiciness 
as abundant or redundant audiovisual feedback triggered by player action \cite{hicks2019juicy}, and frame it as a phenomenon emerging from the coherent integration of mechanics, visuals, and confirmatory feedback \cite{hicks2018good}. 
Kao et al. similarly note that the term was used in different ways, ranging from person-external feedback features to person-internal experiential qualities, 
while recurring themes from literature include moment-to-moment action-contingent response and an amplified (or excessive) amount of feedback relative to input \cite{kao2024does}. 

Moreover, there are more works that have also extended the concept beyond heavy visual embellishments or effects to audio \cite{kao2020effects}, haptic \cite{singhal2021juicy}, and even text-based feedback \cite{fabre2024juicy}. 
These variations make juiciness a productive but ambiguous term. 
Following recent work \cite{kao2024does} in human-computer interaction (HCI), we acknowledge that it is essential for any study engaging with this term to clarify the specific definition or context it adopts, and, thus, here in this study, we use their definition of juiciness as ``\textbf{\textit{the design feature of immediate excessive amounts of feedback in relation to user input at the level of moment-to-moment interaction}}'' \cite{kao2024does}.
The use of this definition does not attempt to cover every quality sometimes associated with juicy design (e.g., haptic or other multisensory aspects), but is based on this operational (and we consider tractable) definition to translate the juiciness concept into visualization interaction.

\subsection{Feedback Design in Visualization Interaction}

Having scoped this study as abovementioned, in this section, we situate our work within prior visualization studies on feedback design. 
In the context of interactive visualization, feedback has been discussed less as an isolated aesthetic layer than as a condition for effective and engaging interaction. 
Elmqvist et al. \cite{elmqvist2011fluid} argue that interaction is central to users' dialogue with data and propose Fluid Interaction as a unifying design concept for information visualization. 
Within this framework, design guidelines such as ``\textit{provide immediate visual feedback on interaction}'', and ``\textit{reward interaction}'' are treated as key properties of the fluidity. 
Their account links feedback not only to responsiveness, but also to sense of control, and the minimization of Norman's \textit{Gulfs of Execution} (the gap between a user's intentions and the system's allowable actions), and \textit{Gulf of Evaluation} (the gap between the system's actual state and the user's perception) \cite{norman1986user}. 
Notably, Elmqvist et al. also 
suggest that well-designed visualization interaction may share similarities with games \cite{elmqvist2011fluid}, an insight our work recalls: we draw on video games' design strategies of the juicy concept to further enhance users' active interaction with data.

Subsequent work has made the role of feedback more explicit. 
Li and Zhu examine visual feedback for interaction in 3D visualization and argue that feedback helps users understand system function, state, and outcome \cite{li2022functional}. 
They further distinguish \textit{feedforward}, \textit{real-time feedback}, and \textit{delay-time feedback}, showing that feedback can support interaction by confirming actions, anticipating possible actions, and clarifying consequences.
This temporal framing is closely aligned with our interest in exploring and specifying juicy feedback design spaces in interactive visualization.
Moreover, other visualization works broaden what feedback can mean and why it matters. 
Chundury et al., for example, show that interaction support need not be purely visual -- sound and touch can function together as feedback channels for understanding chart semantics \cite{chundury2021towards}. 
Willett et al., from a different angle, suggest that exaggerated feedback may shape not only task success but also users' felt sense of agency and capability \cite{willett2021perception}. 

Recent works by Campos \cite{campos2025assessing,campos2025juiciness,campos2025juicy} come closest to 
the present topic, but operate at a different level of analysis.
In one study, Campos surveys 195 publicly accessible interactive visualizations and interprets juiciness through five observable effect categories (animation, particles, audio feedback, screen shake, and persistence), giving an initial descriptive inventory of how such features appear in existing practice 
\cite{campos2025assessing}.
Importantly, this does not suggest that juiciness has been systematically theorized in visualization; rather, it indicates that techniques now legible through a juicy lens already appear in practice, but without a shared conceptual or taxonomy-grounded account.
A separate ethnography, drawing on interviews with designers, suggests that juicy design may enhance engagement, understanding, and memorability, while also risking distraction and overstimulation \cite{campos2025juiciness}. 

Most directly related is Campos's comparative \textit{juicy-versus-dry} study \cite{campos2025juicy}, to our knowledge, one of the few user studies of how juicy design affects engagement and short-term retention in visualization settings; 
it reports no significant difference between the dry and juicy versions, with mixed effects on retention.
We extend this opening from a representation-level has/has-not, juicy-versus-dry\footnote{For the naming of ``\textit{dry},'' i.e., ``\textbf{\textit{non-juicy}},'' hereafter we use non-juicy to unify relevant naming such as ``\textit{dry},'' ``\textit{not juiciness},'' ``\textit{no juicy},'' etc.} \cite{campos2025juicy} contrast to a theory-informed design space: we map juicy feedback onto established interaction categories \cite{yi2007toward}, build controlled stimuli around interaction events, and examine how juiciness varies across interaction type, temporal placement, and intensity. 
Our controlled exemplars and exploratory evaluation do not claim a definitive universal effect, but to characterize how different juiciness dimensions may shape experience across interaction types.
Campos's work helps open the problem space; our study seeks to systematize it through a more explicit taxonomy-grounded operationalization of juicy visualization feedback.

\section{Study Design: Unpacking Design Space Dimensions of Juicy Interactive Visualization}
\label{sec:design}

Grounded in Yi et al.'s taxonomy of visualization interaction \cite{yi2007toward} and Kao et al.'s operational definition of juicy \cite{kao2024does}, we operationalize juicy feedback as a theory-informed design space for interactive visualization. 
We decompose juiciness, often treated as a single stylistic toggle \cite{campos2025juicy}, into a set of analytically distinct and experimentally controllable dimensions. 
This operationalization is an exploratory research instrument that makes an under-specified concept tractable for design and empirical comparison, without claiming to exhaust all forms of juiciness. 
We first lay out a palette of juicy feedback techniques used to build and document our stimuli (\cref{sec:palette}), then unpack three design dimensions against a non-juicy baseline:
\textbf{\textit{juiciness interaction types}}, \textbf{\textit{juiciness feedback types}}, and \textbf{\textit{juiciness intensity-parameter}} (\cref{sec:dimension1,sec:dimension2,sec:dimension3}).
These dimensions are operationally separable, and each of our three studies isolates one while holding the others fixed; we treat possible interactions among them, and further dimensions such as feedback modality, as beyond our exploratory scope (\cref{sec:discussion}).

\subsection{A Palette of Juicy Feedback Techniques}
\label{sec:palette}

Building juicy stimuli systematically first requires a vocabulary of the feedback techniques available to a visualization designer. 
We ground this vocabulary in Campos's content analysis of interactive infographics \cite{campos2025assessing}, which identifies five recurring, observable effect categories: animation, particles, audio feedback, screen shake, and persistence. 
Campos's work codes whether such effects occur, but not how they are realized or tied to an interaction. 
We therefore extend the five categories into a finer set of implementation-grounded sub-categories, the \textit{\textbf{Juice Elements Palette}} (\cref{tab:juicepalette}). 
Each technique in our prototypes is assigned to one sub-category by its dominant mechanism: sound-producing cues to audio feedback, whole-view displacement to screen shake, small spawned units to particles, and visual traces that remain during or after an event (e.g., guides, previews, sustained highlights) to persistence; the remaining motion and transformation effects (e.g., pops, pulses, flashes, sweeps) fall under animation. 
This palette is an operational vocabulary, not an exhaustive taxonomy: every sub-category is derived from a technique we actually instantiated. 
We use it both to construct the stimuli and to document them, with the colored tags in \cref{tab:study1design} and \cref{tab:study2design} tracing each described effect back to its sub-category, making the feedback descriptions inspectable against shared reference.


\renewcommand{\JuiceGroup}[3]{%
  \noalign{\vskip 1.8pt}%
  \multicolumn{3}{%
    @{}%
    >{\columncolor{#1}[0pt][0pt]\color{white}\arraybackslash}%
    m{\columnwidth}%
    @{}%
  }{%
    \begin{tabularx}{\linewidth}{%
      @{}%
      >{\centering\arraybackslash}m{1.85cm}%
      @{\hspace{3pt}}%
      >{\raggedright\arraybackslash}X%
      @{}%
    }
      {\fontsize{7.3pt}{7.9pt}\selectfont
       \sffamily\bfseries #2}
      &
      {\fontsize{6.9pt}{7.5pt}\selectfont
       \sffamily #3}
    \end{tabularx}%
  }%
  \\%
  \noalign{\vskip 1.8pt}%
}

\renewcommand{\JuicePaletteRow}[4]{%
  \rowcolor{#1}%
  \textbf{#2}
  &
  #3
  &
  #4
  \\%
  \arrayrulecolor{white}\hline
}



\begin{table}[!t]

  \setlength{\abovecaptionskip}{1pt}
  \setlength{\belowcaptionskip}{2pt}

  \caption{%
    The \textit{Juice Elements Palette} used to construct and document the JuicyVIS stimuli, extending Campos's five observable effect categories \cite{campos2025assessing} into implementation-grounded sub-categories. Category colors and tag prefixes are reused in \cref{tab:study1design,tab:study2design}. 
    The palette is an operational vocabulary, not an exhaustive taxonomy.%
  }
  \label{tab:juicepalette}

  \centering

  \fontsize{7.9pt}{8.8pt}\selectfont
  \setlength{\tabcolsep}{1.5pt}
  \setlength{\arrayrulewidth}{0.25pt}
  \renewcommand{\arraystretch}{1.13}
  \renewcommand{\tabularxcolumn}[1]{m{#1}}

  \begin{tabularx}{\columnwidth}{%
    @{}
    >{\centering\arraybackslash}m{1.85cm}
    >{\centering\arraybackslash}m{0.90cm}
    >{\raggedright\arraybackslash}X
    @{}%
  }

    \toprule

    \rowcolor{JuiceHeader}
    \textbf{Sub-categories}
      &
    \textbf{Tag}
      &
    \textbf{Mechanism (definition)}
    \\

    \midrule


    \JuiceGroup
      {JuiceAnimDark}
      {ANIMATION}
      {frame transitions creating the illusion of motion; rotations,
       size variations, multi-axis movement, distortions (Campos)}

    \JuicePaletteRow
      {JuiceAnimLight}
      {Pop \& Bounce}
      {\jA{PB}}
      {Mark-level elastic pop, bounce, or press ripple.}

    \JuicePaletteRow
      {JuiceAnimLight}
      {Pulse \& Scale}
      {\jA{PS}}
      {Mark pulsing or enlargement in size or opacity.}

    \JuicePaletteRow
      {JuiceAnimLight}
      {Flash \& Glow}
      {\jA{FG}}
      {A brief flash or glow of luminance.}

    \JuicePaletteRow
      {JuiceAnimLight}
      {Sweep \& Scan}
      {\jA{SS}}
      {A moving highlight band sweeping across the view.}

    \JuicePaletteRow
      {JuiceAnimLight}
      {Transition \& Reveal}
      {\jA{TR}}
      {View- or representation-level transition, entrance/exit,
       staggered reveal, or blur-to-clarity.}


    \JuiceGroup
      {JuiceParticleDark}
      {PARTICLES}
      {small discrete units (single pixels or groups) that spawn on events,
       governed by physics/quantity/speed (Campos)}

    \JuicePaletteRow
      {JuiceParticleLight}
      {Commit Burst}
      {\jPa{CB}}
      {Confetti-like particles bursting on commit.}

    \JuicePaletteRow
      {JuiceParticleLight}
      {Decorative Sparkle}
      {\jPa{DS}}
      {Decorative trailing sparkle particles.}


    \JuiceGroup
      {JuiceAudioDark}
      {AUDIO FEEDBACK}
      {music and sound effects that respond to events (Campos);
       subcategories by interaction phase}

    \JuicePaletteRow
      {JuiceAudioLight}
      {Anticipatory Cue}
      {\jAu{AC}}
      {Soft cue sound on hover or preview, before the action.}

    \JuicePaletteRow
      {JuiceAudioLight}
      {Action / Scrub Sound}
      {\jAu{AS}}
      {Process sound during an ongoing manipulation.}

    \JuicePaletteRow
      {JuiceAudioLight}
      {Completion Sound}
      {\jAu{CS}}
      {Confirming sound on completion or commit.}

    \JuicePaletteRow
      {JuiceAudioLight}
      {Error Sound}
      {\jAu{ES}}
      {Negative sound for an invalid action.}


    \JuiceGroup
      {JuiceShakeDark}
      {SCREEN SHAKE}
      {displacement of the whole screen and its visual elements (Campos)}

    \JuicePaletteRow
      {JuiceShakeLight}
      {Impact Shake}
      {\jS{IS}}
      {Impact shake on commit or positive feedback.}

    \JuicePaletteRow
      {JuiceShakeLight}
      {Error Shake}
      {\jS{ErS}}
      {Negative-feedback shake for an invalid action.}

    \JuicePaletteRow
      {JuiceShakeLight}
      {Sustained Jitter}
      {\jS{SJ}}
      {Low-amplitude tremor sustained during manipulation.}


    \JuiceGroup
      {JuicePersistDark}
      {PERSISTENCE}
      {visual traces left by events; permanent/temporary,
       instant effects/lasting changes (Campos);
       subcategories by trace form}

    \JuicePaletteRow
      {JuicePersistLight}
      {Transient Guides \& Cues}
      {\jP{TG}}
      {Echoing Campos's ``Instant'' persistence; Guides, rings, borders,
       or vignette shown during input or an active state, removed when
       the interaction ends.}

    \JuicePaletteRow
      {JuicePersistLight}
      {Preview / Ghost Overlay}
      {\jP{PG}}
      {Echoing Campos's ``Temporary'' persistence; Semi-transparent
       overlay previewing the potential outcome on hover, dismissed
       on leave.}

    \JuicePaletteRow
      {JuicePersistLight}
      {Sustained Highlight \& Dim}
      {\jP{SH}}
      {Echoing Campos's ``Lasting'' persistence; Highlight, tint, or
       de-emphasis held while a selection, filter, or mode is active.}

    \JuicePaletteRow
      {JuicePersistLight}
      {Status Badge \& Stamp}
      {\jP{SB}}
      {Held after the action, until the next action; Discrete status
       marker that appears on completion and persists.}

    \JuicePaletteRow
      {JuicePersistLight}
      {Persisted Trace \& Frame}
      {\jP{PT}}
      {Echoing Campos's ``Lasting'' persistence; Line, frame, or mark
       left after the event, holding or fading.}

    \arrayrulecolor{black}
    \bottomrule

  \end{tabularx}
  \vspace{-15pt}
\end{table}


To instantiate this design space, we developed \href{https://juicy-vis-10aac.web.app/}{\textbf{\textit{JuicyVIS\(\nearrow\)}}} (\href{https://github.com/revisit-games/juicyvis}{\textit{GitHub} repo\(\nearrow\)}), a web-based prototype (\textit{React}, \textit{TypeScript}, \textit{D3.js}) in which condition-specific infographic variants (hereafter, ``\textit{infographic}'') are served from a shared codebase and parameterized through URL query strings;
this implementation underlies the interactive stimuli linked in the following subsections.
The stimuli were constructed from publicly accessible datasets obtained through \textit{Our World in Data} (\textit{OWID}) \cite{OWID2025EpochAI,OWID2025ChildMortality,OWID2025Energy,OWID2025MeatandFish,OWID2024Age,OWID2025Gini,OWID2026Internet,OWID2026Electricity,OWID2025Cooking,OWID2026Renewable,OWID2026Urban,OWID2025Immunization,OWID2024Smoke,OWID2025Obesity}.
We followed the source attribution and citation guidance provided on the corresponding OWID pages when reconstructing custom study stimuli from the data.

\subsection{Dimension 1: Juiciness Interaction-Types}
\label{sec:dimension1}

The first dimension concerns where juiciness is attached in the interaction types. 
We operationalize it at the level of Yi et al.'s seven interaction categories \cite{yi2007toward}, choosing these seven because together they cover that taxonomy's full intent space rather than an ad hoc subset.
JuicyVIS includes an infographic for each category (Select, Explore, Reconfigure, Encode, Abstract/Elaborate\footnote{For ``Abstract/Elaborate'', hereafter we use ``\textit{Abstract}'' for short.}, Filter, and Connect); for each chart, we built a matched pair, a non-juicy baseline and a juicy variant, yielding \textbf{14} conditions (for Dimension 1). 
The juicy techniques attached to each interaction are drawn from the palette (\cref{sec:palette}) and chosen to fit that interaction's affordances, for instance a commit-and-confirm pattern for Reconfigure's sorting action and matched-range cues for Filter's thresholding; \cref{tab:study1design} documents the full mapping and links to the prototypes.


\begin{table}[!t]
  \renewcommand{\arraystretch}{1.20}
  \renewcommand{\tabularxcolumn}[1]{m{#1}}
  \setlength{\tabcolsep}{2.2pt}

  \caption{%
    Juicy feedback implemented for each interaction type \cite{yi2007toward}, with links to the anonymized juicy and non-juicy prototypes. Colored terms and tags map each effect to a sub-category of the \textit{Juice Elements Palette} \cref{tab:juicepalette}. Because juicy feedback is multi-aspected and interaction-contingent, below are compact summaries; we encourage readers to inspect the linked prototypes directly.%
  }
  \label{tab:study1design}

  \scriptsize
  \centering

  \begin{tabularx}{\columnwidth}{%
    @{}
    >{\columncolor[gray]{0.94}\centering\arraybackslash}m{1.35cm}
    >{\raggedright\arraybackslash}X
    >{\columncolor[gray]{0.94}\raggedright\arraybackslash}m{1.65cm}
    @{}%
  }

    \toprule
    \textbf{Interaction Types}
      & \textbf{Juicy Feedback Descriptions}
      & \textbf{Prototype Links}
    \\
    \midrule

    \textbf{Select}
    &
    Adds \jPTerm{crosshair guides}{TG};
    hover adds \jAuTerm{sound}{AC},
    \jATerm{point pop}{PB},
    \jPTerm{category tint}{SH}, and
    \jPTerm{de-emphasis}{SH};
    click adds \jAuTerm{confirmation sound}{CS},
    \jSTerm{shake}{IS},
    \jPaTerm{particles}{CB}, and
    \jATerm{pop highlight}{PB}.
    &
    \StudyOnePrototypeLinks{1}
    \\
    \hdashline

    \textbf{Explore}
    &
    Adds pan-zoom grab cues
    (\jPTerm{press ring}{TG},
    \jPTerm{vignette}{TG},
    \jPTerm{border}{TG},
    \jPTerm{crosshair}{TG},
    \jPTerm{arrows}{TG});
    panning adds \jAuTerm{motion sounds}{AS} and
    \jATerm{arrow flashes}{FG};
    zooming adds \jAuTerm{stepwise sounds}{AS}.
    &
    \StudyOnePrototypeLinks{2}
    \\
    \hdashline

    \textbf{Reconfigure}
    &
    Adds \jPTerm{slider arrows}{TG} and a
    \jPTerm{translucent reorder preview}{PG};
    commit triggers
    \jATerm{sorting animation}{TR},
    \jAuTerm{tick sounds}{AS},
    \jAuTerm{confirmation cue}{CS},
    \jSTerm{shake}{IS}, and
    \jPaTerm{particles}{CB}.
    &
    \StudyOnePrototypeLinks{3}
    \\
    \hdashline

    \textbf{Encode}
    &
    Adds
    \jPTerm{hover previews of alter representations}{PG};
    switches use
    \jATerm{animated transitions}{TR} with
    \jPTerm{vignette}{TG},
    \jPTerm{outline}{TG},
    \jSTerm{shake}{IS},
    staged \jATerm{motion}{TR}/\jAuTerm{sound}{AS}, and
    \jAuTerm{completion cues}{CS};
    color changes add
    \jATerm{smooth recoloring}{TR} and
    \jPTerm{local highlight}{SH}.
    &
    \StudyOnePrototypeLinks{4}
    \\
    \hdashline

    \textbf{Abstract}
    &
    Adds drill feedback via
    \jATerm{animated tile entry/exit}{TR},
    \jAuTerm{transition audio}{AS},
    \jATerm{brief blur/frame overlays}{TR}, and
    \jPaTerm{particles}{CB};
    invalid drills trigger an
    \jAuTerm{error cue}{ES} and
    \jSTerm{shake}{ErS}.
    &
    \StudyOnePrototypeLinks{5}
    \\
    \hdashline

    \textbf{Filter}
    &
    Adds slider-hover cues via
    \jPTerm{vignette}{TG},
    \jPTerm{outline}{TG}, and
    \jPTerm{threshold highlighting}{TG};
    scrubbing adds
    \jAuTerm{tick sounds}{AS},
    \jATerm{flashes}{FG}, and
    \jATerm{animated line entry/exit}{TR};
    completion adds
    \jSTerm{shake}{IS} and a
    \jATerm{popping}{PB}
    \jPTerm{matched-count badge}{SB}.
    &
    \StudyOnePrototypeLinks{6}
    \\
    \hdashline

    \textbf{Connect}
    &
    Adds cross-view hover linking via
    \jPTerm{hovered-mark emphasis}{SH} and an
    \jATerm{animated bridge cue}{TR};
    selection adds
    \jAuTerm{connecting sound}{CS},
    \jATerm{impact flash}{FG},
    \jPTerm{selected-ring animation}{SH},
    \jSTerm{shake}{IS}, and
    \jATerm{staggered reveal of related items}{TR}.
    &
    \StudyOnePrototypeLinks{7}
    \\

    \bottomrule
  \end{tabularx}
  \vspace{-14pt}
\end{table}



We do not frame our implementations as a universal recipe. 
The effects and appropriateness of juicy feedback are contingent on interaction intent, representational context, and timing, and a feedback family that helps one interaction may play a different role, or even become inappropriate in another. 
Our aim is controlled, theory-grounded variation: a principled set of exemplars that makes cross-interaction comparison possible. 
These exemplars were designed by the authors from this theoretical grounding and refined through pilot tests; they were not validated by external design professionals, a step we leave to future work (\cref{sec:discussion}).

\subsection{Dimension 2: Juiciness Feedback-Types}
\label{sec:dimension2}

Beyond which interaction type is juicified (Dimension 1), a second dimension concerns when juicy feedback is presented relative to the unfolding interaction.
Prior visualization work has shown that feedback can be temporally differentiated
\cite{li2022functional}; we build on this insight for a different purpose, 
parameterizing temporal placement as a design space dimension of juicy visualization feedback.
We operationalize three compact feedback types, \textbf{\textit{Pre-interaction}}, \textbf{\textit{In-interaction}}, and \textbf{\textit{Post-interaction}}, named for 
when juicy cues are attached relative to a focal interaction event in our stimuli.


\begin{table*}[!t]

  \renewcommand{\arraystretch}{1.14}
  \setlength{\tabcolsep}{2pt}

  \caption{%
    \textit{Pre-}, \textit{in-}, and \textit{post-interaction} juicy
    feedback for each interaction type; colored
    terms and tags map each effect to a sub-category of the \textit{Juice Elements Palette} (\cref{tab:juicepalette}). 
    Prototype links:
    \StudyTwoPrototypeLink{0}{Non-juicy Baseline} |
    \StudyTwoPrototypeLink{1}{Pre-Interaction} |
    \StudyTwoPrototypeLink{2}{In-Interaction} |
    \StudyTwoPrototypeLink{3}{Post-Interaction} |
    \StudyTwoPrototypeLink{4}{Pre$+$In-Interaction} |
    \StudyTwoPrototypeLink{5}{In$+$Post-Interaction} |
    \StudyTwoPrototypeLink{6}{Pre$+$Post-Interaction} |
    \StudyTwoPrototypeLink{7}{Pre$+$In$+$Post-Interaction}.%
  }
  \label{tab:study2design}

  \scriptsize
  \centering

  \begin{tabular}{%
    @{}
    >{\centering\arraybackslash}m{1.30cm}
    @{\hspace{5pt}}
    >{\columncolor[gray]{0.94}[0pt][0pt]
      \raggedright\arraybackslash}m{4.65cm}
    @{\hspace{5pt}}
    >{\raggedright\arraybackslash}m{5.05cm}
    @{\hspace{5pt}}
    >{\columncolor[gray]{0.94}[0pt][0pt]
      \raggedright\arraybackslash}m{5.95cm}
    @{}%
  }

    \toprule

    \textbf{Types}
      &
    \textbf{Pre-Interaction Juicy Feedback}
      &
    \textbf{In-Interaction Juicy Feedback}
      &
    \textbf{Post-Interaction Juicy Feedback}
    \\

    \midrule


    \textbf{Select}
    &
    Cursor-following guides
    (\jPTerm{crosshair and horizontal line}{TG});
    non-committal
    \jATerm{press ripple}{PB} and
    \jAuTerm{sound}{AC};
    \jPTerm{hover highlight}{SH} and
    \jATerm{mark enlargement}{PS}.
    &
    Click commit triggers a
    \jAuTerm{pop sound}{AS},
    \jATerm{impact flash}{FG}, and
    \jPTerm{cursor ring}{TG};
    selected and related marks
    \jATerm{briefly glow}{FG}.
    &
    After settling, a
    \jATerm{title pulse}{PS} and
    \jPTerm{status badge}{SB} appear with a
    \jAuTerm{delayed chime}{CS};
    Views show
    \jPaTerm{fireworks}{CB},
    \jPTerm{cascading red frames}{PT}, and a
    \jATerm{delayed bar bounce}{PB};
    selected marks retain a
    \jPTerm{subtle blink until reselected}{SH}.
    \\

    \hdashline


    \textbf{Explore}
    &
    Hovering year controls triggers
    \jPTerm{recolored blinking focus or window guides}{TG},
    \jPTerm{arrow cues}{TG}, and
    \jAuTerm{sound}{AC}.
    &
    Scrubbing triggers
    \jAuTerm{tick sounds}{AS},
    \jPTerm{vignette}{TG},
    \jPTerm{teal active-slider highlight}{SH}, and
    \jPTerm{pulsing focus guides}{TG}.
    &
    On release,
    \jATerm{window guides settle}{TR},
    followed by a prominent
    \jATerm{red sweep}{SS},
    \jATerm{flash}{FG},
    \jPTerm{updated badge}{SB}, and
    \jAuTerm{completion chime}{CS}.
    \\

    \hdashline


    \textbf{Reconfigure}
    &
    Hovering reconfigure options reveals a
    \jPTerm{ghost preview with distinct reordered styling}{PG},
    \jPTerm{baseline dimming}{SH}, and
    \jAuTerm{sound}{AC}.
    &
    Sorting interaction triggers
    \jAuTerm{pop sounds}{AS} and a
    \jATerm{color-shifted button burst}{FG};
    the view \jATerm{reorders}{TR} with
    \jATerm{bar pop-in}{PB} and a
    \jATerm{header scan}{SS}.
    &
    After sorting, a
    \jPTerm{sorted badge}{SB} appears with a
    \jATerm{header confirmation scan}{SS};
    a delayed
    \jATerm{top-down bar bounce}{PB} and
    \jPTerm{red frame highlight}{PT} follow,
    ending with a \jAuTerm{chime}{CS}.
    \\

    \hdashline


    \textbf{Encode}
    &
    Hovering encode options reveals a
    \jPTerm{recolored ghost alt-representation overlay}{PG},
    \jPTerm{mini glyph hint}{PG}, and
    \jAuTerm{sound effect}{AC}.
    &
    Toggle encoding plays
    \jAuTerm{pop sounds}{AS},
    \jATerm{brief blur/saturate}{TR}, and
    \jATerm{subtle scale pulse}{PS};
    The view \jATerm{crossfades}{TR} while
    \jATerm{staggered mark pop-in}{PB}.
    &
    After encoding, a
    \jPTerm{status badge}{SB},
    \jATerm{clarity-settle overlay}{TR}, and
    \jAuTerm{chime}{CS} appear;
    the histogram performs repeated
    \jATerm{bar bounces}{PB} with
    \jPTerm{highlights}{SH},
    while the beeswarm shows repeated
    \jPaTerm{sparkle sweeps}{DS}.
    \\

    \hdashline


    \textbf{Abstract}
    &
    Hovering abstract options shows a
    \jPTerm{tooltip or preview card}{PG},
    \jPTerm{anchor highlight}{SH}, and
    \jAuTerm{sound}{AC}.
    &
    Clicking the detail option triggers
    \jAuTerm{pop sounds}{AS} and, for
    \textit{Detail} 2, a
    \jATerm{top-down staggered reveal of per-row metrics through motion and opacity}{TR}.
    &
    After the detail change, a
    \jPTerm{status badge}{SB} appears,
    followed by a delayed
    \jATerm{cascade flash of metrics and year ticks in red bold styling}{FG},
    then a \jAuTerm{chime}{CS} and
    \jATerm{restore}{TR}.
    \\

    \hdashline


    \textbf{Filter}
    &
    Hovering slider surfaces triggers
    \jPTerm{blinking and highlighted range guides}{TG};
    hovering region legend chips shows a
    \jPTerm{recolored group outline}{SH} while
    \jPTerm{others dim}{SH}.
    &
    Dragging sliders triggers
    \jPTerm{scrubbing vignette}{TG},
    \jPTerm{teal scrub highlights}{SH},
    \jPTerm{emphasized guides}{TG}, and
    \jATerm{smooth de-emphasis transitions}{TR};
    region toggling adds
    \jAuTerm{pop sounds}{AS} and
    \jSTerm{subtle view jitter}{SJ}.
    &
    After filtering, an indicator
    \jATerm{pulses}{PS} with a
    \jPTerm{stamp}{SB};
    \jATerm{red sweeps}{SS} confirm updates across views,
    followed by a \jAuTerm{chime}{CS};
    the filtering legend also shows a
    \jPTerm{delayed red outline}{PT} with
    \jATerm{flash effects}{FG}.
    \\

    \hdashline


    \textbf{Connect}
    &
    Hovering any items draws a
    \jPTerm{dashed bridge}{PT} linking views with the
    \jPTerm{linked mark emphasis}{SH}.
    &
    On selection, drawing a
    \jPTerm{dashed bridge line(s)}{PT}
    to show connections, then
    \jPTerm{holds/fades}{PT}.
    &
    After related items are
    \jATerm{revealed}{TR} and the
    \jPTerm{bridge line(s) drawn}{PT},
    the line(s) persisted with a
    \jPTerm{moving dashed effect}{PT}\,
    \jAInline{TR}
    and
    \jPTerm{blinking}{PT}.
    \\

    \bottomrule

  \end{tabular}
  \vspace{-12pt}
\end{table*}


We further motivate these categories through Hicks et al.’s account of juicy direct feedback \cite{hicks2018good}. 
In our reading, pre-interaction juicy feedback echoes the ``\textit{Confirmatory}'' feedback, ``a direct response to physical input of a button,'' while in-interaction echoes their ``\textit{Unambiguous}'' feedback, ``information [that can] be connected to actions and only interpreted in one way'' \cite{hicks2018good}. 
Mapped onto interactive visualization, this yields a useful distinction: \textit{confirmatory} feedback responds to physical input even when it does not trigger a functional interaction (e.g., clicking on blank space, pressing an unbound key, or hovering over a button without clicking), 
whereas \textit{unambiguous} feedback  is tightly coupled to the execution of an interaction on an interface element (e.g., the moment when a bound button is being pressed). 
Post-interaction feedback aligns with their ``\textit{Supplementary feedback}'' \cite{hicks2018good}, reinforcing the perceived consequence after the primary interaction completes (e.g., a completion cue following a successful filter change). 

These categories are not exhaustive; 
they offer a compact, theory-informed way of varying the temporal placement of juicy feedback across the seven interaction types from Dimension 1. 
\cref{tab:study2design} summarizes the pre-, in-, and post-interaction feedback implemented for each type.
Combined with a non-juicy baseline, the three feedback types were instantiated as seven juicy combinations (\textit{pre-only}, \textit{in-only}, \textit{post-only}, \textit{pre$+$in}, \textit{in$+$post}, \textit{pre$+$post}, and \textit{pre$+$in$+$post}), yielded \textbf{8} conditions (for Dimension 2).

\subsection{Dimension 3: Juiciness Intensity-Parameter}
\label{sec:dimension3}

A third dimension concerns how strongly juicy feedback is expressed. 
Prior game studies have operationalized juiciness as graded intensity rather than mere presence: 
Kao compares four juiciness levels (\textit{none}, \textit{medium}, \textit{high}, and \textit{extreme}) and shows that both insufficient and excessive juiciness can be detrimental, so degree matters as much as presence \cite{kao2020effects}; 
Singhal and Schneider manage \textit{low} and \textit{high} haptic embellishment, manipulating intensity
as a design variable \cite{singhal2021juicy}.
These levels, however, have typically been set through designer judgment and iteration rather than a reported, reproducible parameterization \cite{kao2024does}; we make ours explicit.

We operationalize juicy intensity as a parameterized scaling of the feedback techniques from Dimensions 1 and 2. 
Four parameters are varied around the medium preset, which is our as-designed juicy version and serves as the \(1.00\times\) reference: 
motion and deformation amplitude, effect duration, audio-cue volume, and transient visual-effect density; the effective particle count is derived from density through a mild super-linear mapping ($m_{\mathrm{VFX}}^{1.35}$), so that denser bursts remain perceptually separable. 
We began from a symmetric bracket around the reference and tuned each parameter's low and high multipliers in a small pilot, so that the three levels were reliably distinguishable while medium stayed the designed default; 
the final multipliers are specified independently per parameter rather than as one uniform scale (\cref{tab:study3design}). 
With a non-juicy baseline that disables these effects, we obtain four intensity conditions: 
\href{https://juicy-vis-10aac.web.app/?chart=9&juicy=0}{\textbf{Non-juicy\(\nearrow\)}}, 
\href{https://juicy-vis-10aac.web.app/?chart=9&juicy=1}{\textbf{Low\(\nearrow\)}}, 
\href{https://juicy-vis-10aac.web.app/?chart=9&juicy=2}{\textbf{Medium\(\nearrow\)}}, and 
\href{https://juicy-vis-10aac.web.app/?chart=9&juicy=3}{\textbf{High\(\nearrow\)}}. 
These presets are exemplar settings, offered as a transparent, reproducible reference for future work; we make no claim that they are optimal or unique. They let us examine whether juicy visualization feedback exhibits not only categorical but also ``\textit{dose}''-dependent effects.

\begin{table}[t]
\caption{Parameter preset bundles operationalizing low, medium, and high juiciness intensity in Study 3. Multipliers are specified independently per parameter rather than as a uniform linear scale; the medium preset is our as-designed juicy version and serves as the \(1.00\times\) reference. Effective particle count is computed as \(m_{\mathrm{VFX}}^{1.35}\) before rounding and implementation-specific caps. The non-juicy condition disables these effects.}
\vspace{-6pt}
\label{tab:study3design}

\centering
\scriptsize
\setlength{\tabcolsep}{4pt}
\renewcommand{\arraystretch}{1.08}

\begin{threeparttable}
\begin{tabular*}{\columnwidth}{@{\extracolsep{\fill}}
>{\columncolor[gray]{0.93}[2pt][2pt]
  \raggedright\arraybackslash}m{0.52\columnwidth}
>{\centering\arraybackslash}m{0.115\columnwidth}
>{\columncolor[gray]{0.93}[2pt][2pt]
  \centering\arraybackslash}m{0.115\columnwidth}
>{\centering\arraybackslash}m{0.115\columnwidth}
@{}}

\toprule

\parbox[c][2.2ex][c]{\linewidth}{%
  \raggedright\textbf{Parameter dimension}%
}
&
\parbox[c][2.2ex][c]{\linewidth}{%
  \centering\textbf{Low}\\
}
&
\parbox[c][2.2ex][c]{\linewidth}{%
  \centering\textbf{Medium}\\
}
&
\parbox[c][2.2ex][c]{\linewidth}{%
  \centering\textbf{High}\\
}
\\

\midrule

Motion/deformation amplitude
(\(m_{\mathrm{A}}\))
&
\(0.10\times\)
&
\(1.00\times\)
&
\(1.90\times\)
\\

Effect duration
(\(m_{\mathrm{T}}\))
&
\(0.55\times\)
&
\(1.00\times\)
&
\(1.55\times\)
\\

Audio-cue volume
(\(m_{\mathrm{audio}}\))
&
\(0.35\times\)
&
\(1.00\times\)
&
\(1.35\times\)
\\

Transient visual-effect density
(\(m_{\mathrm{VFX}}\))
&
\(0.40\times\)
&
\(1.00\times\)
&
\(2.60\times\)
\\

Effective particle count
(\(m_{\mathrm{VFX}}^{1.35}\))
&
\(0.2903\times\)
&
\(1.0000\times\)
&
\(3.6326\times\)
\\

\bottomrule
\end{tabular*}
\end{threeparttable}
\vspace{-12pt}
\end{table}

\section{Study Methods}

To probe these dimensions empirically, we organized the 26 JuicyVIS prototypes into three online studies (\href{https://github.com/revisit-games/JuicyVIS-studies}{\textit{GitHub} repo\(\nearrow\)}), one per design-space dimension. 
Each study used an exploratory mixed-methods approach: trial-level ratings and task responses provide comparative quantitative evidence on engagement, aesthetic experience, preference, and task outcomes, while open-ended responses captured how participants interpreted juicy feedback. 
We analyzed the open-ended data with \textit{reflexive thematic analysis} (\textit{RTA}) \cite{Braun2019reflexive,Braun2022thematicanalysis}.
All studies were conducted by using \href{https://revisit.dev/}{\textit{reVISit}\(\nearrow\)} \cite{cutler2026revisit2,ding2023revisit,wilburn2025running}, 
with responses stored in \href{https://firebase.google.com/}{\textit{Firebase}\(\nearrow\)}. 
The study were approved by the Institutional Review Board (IRB) at Worcester Polytechnic Institute (Record Number: IRB-26-0441); participants gave informed consent, were briefed on data handling, were de-identified (quoted as P\#), and could withdraw at any time without consequence.

\subsection{Procedure of Each Study}

Each study consists of four sections: 
(i) instructions 
and a digital consent form; 
(ii) an attention check, an audio speaker test (juicy feedback includes sound), 
and a color blindness test (juicy elements rely on color); 
(iii) a series of JuicyVIS trials with tasks and measures (\cref{sec:measures}); and 
(iv) a follow-up questionnaire on overall experience and demographics. 
To prevent learning or carryover effects, each study used a separate participant sample. 
Both Study 1 and Study 2 took about 20-30 minutes, while Study 3 took about 15-20 minutes.
Several conventions were shared across studies: 
Each trial embedded one visualization with three task questions, included both to ensure active interaction and to assess readability and task performance. 
Different trials used different dataset topics to limit repeated exposure to identical content, and trial order was counterbalanced with a Latin-square scheme. 
Within each study, the first encountered trial(s) served as tutorial familiarization and were excluded from analysis.

\subsubsection{Study 1: Juiciness in Interaction Types}

Study 1 (Dimension 1, \cref{sec:dimension1}) examines how juicy vs. non-juicy feedback is experienced across the seven interaction types. 
Each participant completed seven trials, one per type. 
Following the matched-pair sequencing of prior work \cite{bateman2010useful}, 
a participant saw only one variant 
of each interaction type (either the juicy or the non-juicy version; \cref{tab:study1design}), keeping the comparison between-participants at the level of each type while allowing an individual's sequence to mix 
juicy and non-juicy trials; which variant each participant saw was counterbalanced across participants.  
The two versions of a type used identical task questions.
The first two trials were tutorials, leaving five analyzed trials per participant.

\subsubsection{Study 2: Juiciness in Feedback Types}

Study 2 (Dimension 2, \cref{sec:dimension2}) examines how different temporal placements of juicy feedback shape experience. 
Presenting all eight timing conditions in one session would risk fatigue (and therefore have potential fatigue effects), so we adopted an incomplete-block within-subjects design: each participant saw five trials, the non-juicy baseline and the full pre$+$in$+$post condition, plus three of the six intermediate combinations (\cref{tab:study2design}). 
The intermediate conditions were counterbalanced so that all six appeared evenly across the sample.
After the one tutorial trial, four trials per participant were analyzed.

\subsubsection{Study 3: Juiciness with Different Intensity}

Study 3 (Dimension 3, \cref{sec:dimension3}) isolates how intensity relates to experience.
Holding the interaction set and feedback timing fixed, we varied only the intensity preset, giving four conditions. 
In this within-subjects design, each participant completed four trials, one per intensity.
After the one tutorial trial, three trials per participant were analyzed, yielding an incomplete repeated-measures set in which the omitted intensity varied across participants by counterbalanced order.

\subsection{Tasks and Measures}
\label{sec:measures}

Each trial included three task questions for assessing readability and task performance across conditions. 
After the task, participants rated the trial on two scales: 
(i) the \textit{BeauVis} 5-item scale to assess participants' aesthetic experience of the visualization \cite{he2022beauvis}; and 
(ii), to align with prior work on juicy visualization infographics, we used Campos's adapted \textit{User Engagement Scale} Short Form (\textit{UES-SF})  \cite{campos2025juicy}, which was adapted from O'Brien et al.'s work \cite{o2018practical}, to measure engagement in the context of juicy visualization experiences.
After completing all trials within a study, participants answered a set of overall preference questions: a forced-choice default-style preference, an open-ended prompt on which infographic they preferred overall and why, a 10-point slider on their overall preference for adding versus not adding juicy feedback, and a final open-ended prompt explaining their choices.
So that participants could refer to conditions without our seeing the numbering logic, we labeled each condition by its infographic's title-takeaway (e.g., ``\textit{Energy}'' from the ``\textit{Per Capita Energy Consumption by Source}'' infographics). 
In addition, Study 2 asked about preferred feedback timing 
and Study 3 asked about preferred intensity. 

\subsection{Participants}
\label{sec:participants}

We set recruitment targets using a priori simulation-based power analyses in \href{https://www.r-project.org/}{\textit{R\(\nearrow\)}} with linear mixed-effects models, using UES as the primary outcome and $\alpha=.05$. 
For Study 1, we modeled overall juicy versus non-juicy effects as: \verb!UES ~ Juicy + interaction_type + (1 | participant)!, assuming participant-level $SD=0.70$ and residual $SD=0.90$; under a medium assumed juicy effect of 0.25 UES units, the required sample exceeded 60, whereas a large effect of 0.35 suggested about 32. 
We therefore adopted a pragmatic compromise target of 48 completed participants. 
For Study 2, after revising the primary quantitative contrast to a coverage-count trend, we modeled: \verb!UES ~ coverage_num + (1 | participant)!, where coverage ranged from 0 to 3 phases (non-juicy baseline, single type of feedback, double types of feedback, and all) and $participant-level/residual$ $SD$s were set to $0.70/0.80$. 
Under a medium assumed slope of 0.12 UES units per additional covered phase, approximately 72 participants were required, whereas a large slope of 0.18 suggested about 32; because Study 2 is interpreted more exploratory, we set its target to 48. 
For Study 3, we modeled: \verb!UES ~ intensity_num + (1 | participant)! with participant-level / residual $SD$s of 0.70 / 0.80; a medium linear intensity slope of 0.20 yielded a required sample of 28. 
Accordingly, we targeted participants for Study 1 ($N=48$), Study 2 ($N=48$), and Study 3 ($N=28$).
These targets were chosen to support informative comparison across conditions in a broad design space under practical recruitment constraints. 
Accordingly, we interpret the quantitative component of the project as theory-guided comparative evidence rather than as a fully powered confirmatory test of small universal effects.

\begin{figure}[t]
  \centering
  \begin{subfigure}[b]{0.75\columnwidth}
  	\centering
  	\includegraphics[width=\textwidth]{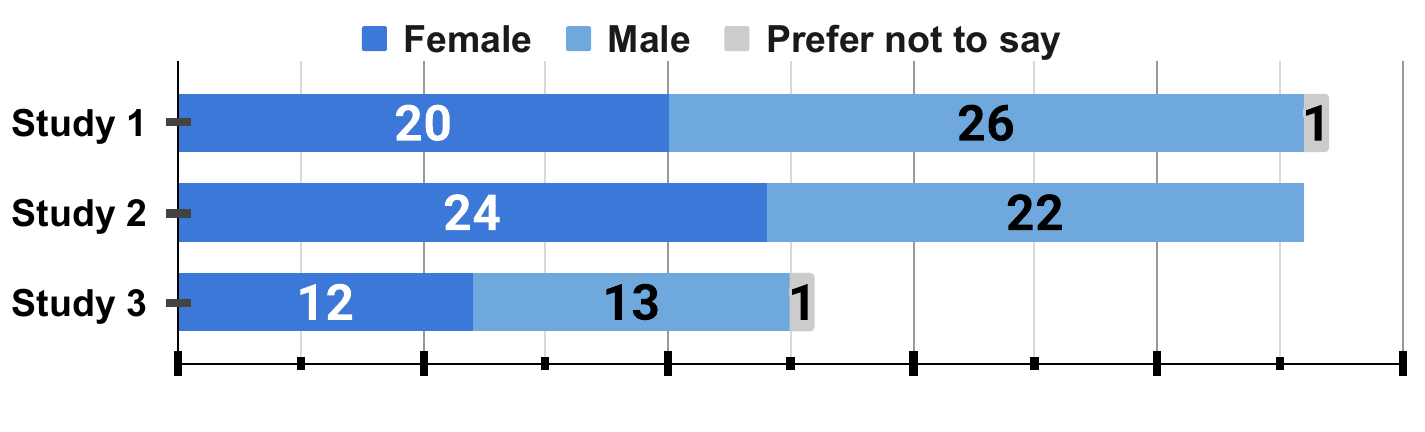}
  	\caption{Gender identity composition}
  	\label{fig:demographics_1}
  \end{subfigure}%
  \hfill%
  \\%
\vspace{-2pt}
  \begin{subfigure}[b]{0.75\columnwidth}
  	\centering
  	\includegraphics[width=\textwidth]{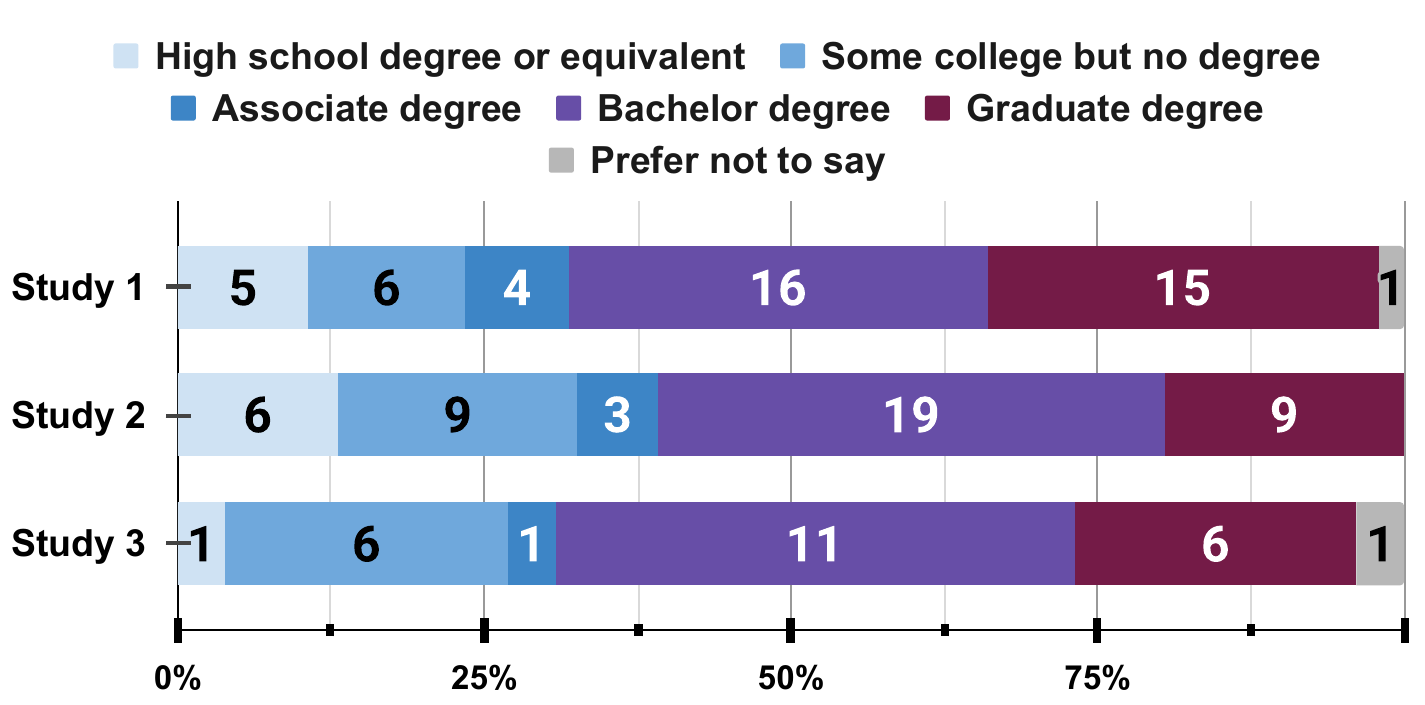}
  	\caption{Educational background composition}
  	\label{fig:demographics_2}
  \end{subfigure}%
  \subfigsCaption{Demographic composition of participants across the three studies, including participants' self-identified gender identity composition (\textbf{a}), and educational background composition (\textbf{b}).}
  \label{fig:demographics}
\vspace{-12pt}
\end{figure}

We recruited from \textit{Prolific} and received 124 submissions, excluding $n=5$ ($4.03\%$) for insufficient responses, and leaving 47, 46, and 26 valid participants for Studies 1-3. The average age of Study 1 was 38.4, 
Study 2 was 36.4, 
and Study 3 was 39.8. 
Participants' self-identified gender identities and educational background are represented in \cref{fig:demographics}.

\subsection{Data Analysis}
\label{sec:dataanalysis}

Across studies, quantitative analyses focused on three primary outcomes: engagement (UES-SF), aesthetic experience (BeauVis), and task performance.
Studies 1, 2, and 3 contributed 235, 184, and 78 analyzed trial records, respectively. 
We use \href{https://www.ibm.com/products/spss-statistics}{\textit{SPSS\(\nearrow\)}} to conduct the analysis.
For each retained trial, task performance was scored as the sum of the three task items, BeauVis as the mean of its five items, and UES-SF following recommended scoring procedures: we computed the four subscale means, 
\textit{Focused Attention} (FA), \textit{Perceived Usability} (PU), \textit{Aesthetic Appeal} (AE), and \textit{Reward Factor} (RW), and took the overall UES-SF as the mean of all items. 
Internal consistency was high across studies (all $\alpha\geq.92$). 
Inferential models matched each study's repeated-measures structure, 
using linear mixed-effects models with participant as a random intercept. 
In Study 1, fixed effects were juiciness condition (juicy vs.\ non-juicy) and interaction type. As a secondary analysis of Study 1, we fit parallel models for the four UES-SF subscales to examine which dimensions of engagement were most affected by juiciness. 
In Study 2, the primary model tested whether broader temporal coverage of juicy feedback predicted outcomes using coverage count (\cref{sec:participants}), and secondary additive models replaced coverage count with binary indicators for pre-, in-, and post-interaction feedback timing. 
In Study 3, primary models tested ordered intensity effects using an ordinal predictor (none, low, medium, high), and secondary follow-up models treated intensity as a four-level factor to probe potential non-monotonic differences across presets. 
Estimated marginal means and Bonferroni-adjusted pairwise comparisons were reported for focal contrasts. 
Because UES-SF is explicitly multidimensional, subscale analyses were treated as secondary and explanatory rather than as substitutes for the total score. 

For the open-ended responses, we used RTA to conduct analysis, which treats the researcher's interpretation as central to coding 
\cite{Braun2019reflexive,Braun2022thematicanalysis}. 
The first two authors first read through the data and used their own language to take notes, then developed codes. 
Specifically, we inductively identified participants' accounts of experiences (no matter positive or negative) through their descriptions, such as ``\textit{easy to use}," ``\textit{distracting},'' ``\textit{annoying},'' ``\textit{fun incentive to explore},'' ``\textit{sense of achievement},'' which we treated as codes. 
We then met to review our codes, examining these codes and descriptions' alignment with each trial condition (that participants were referring to), and collating codes into themes.


\section{Findings}

Our findings are based on a total of 119 submissions (removing 5, or 4.03\%), from Studies 1 ($n=47$), 2 ($n=46$), and 3 ($n=26$), and are organized as follows: results from juiciness in interaction types (Dimension 1; Study 1), juiciness in feedback types (Dimension 2; Study 2), and juiciness in different intensity (Dimension 3; Study 3).
For study transparency and repeatability, we provide de-identified trial-level datasets for all three studies, SPSS analysis syntax, and corresponding output files at the anonymized \href{https://osf.io/bs4cz/overview?view_only=a6151e25b7a34a37b27ccf1199b8d840}{view-only OSF repository\(\nearrow\)}.

\subsection{Evaluating Juiciness in Interaction Types (Study 1)}

Across all retained Study 1 trials ($N=235$, from 47 participants), mean scores were $M=4.74$ for BeauVis total (1-7), $M=3.45$ for UES-SF total (1-5), and $M=2.76$ for task performance (0-3). 
These descriptive results indicated generally positive evaluations overall, with room for condition differences.
Our primary mixed-model analyses (\cref{tab:study1_results}) showed a consistent advantage for juicy over non-juicy trials. 
For overall engagement, juiciness had a significant main effect on UES-SF total (\(p=6.871e-5\)), with higher estimated marginal means for juicy trials ($M=3.647$) than for non-juicy trials ($M=3.224$). 
For aesthetic pleasure, juiciness also had a significant main effect on BeauVis total (\(p=1.809e-6\)), again favoring juicy trials ($M=5.155$) over non-juicy trials ($M=4.284$). 
Finally, task performance showed a smaller but still significant advantage for juicy trials (\(p=1.577e-2\)), with estimated marginal means of $M=2.832$ for juicy trials and $M=2.703$ for non-juicy trials. 
To better understand which aspects of engagement drove the overall UES-SF effect, we fit parallel mixed models for the four UES-SF subscales. 
Juiciness significantly increased Focused Attention, Perceived Usability, Aesthetic Appeal, and Reward (\cref{tab:study1_results}). 

\begin{table}[t]
\caption{Study 1 mixed-model results for juicy versus non-juicy trials. Linear mixed-effects models were fit at the trial level with participant as a random intercept and juiciness condition and interaction type as fixed effects. EMM\(^-\) and EMM\(^+\) denote estimated marginal means for non-juicy and juicy trials, respectively. Interaction type also showed significant main effects across all models (all \(p \le .019\)). Notably, denominator degrees of freedom were estimated using the Satterthwaite approximation, and therefore may vary across tests. Cells shaded in blue indicate statistically significant differences.}
\label{tab:study1_results}
\centering
\scriptsize
\setlength{\tabcolsep}{4pt}
\renewcommand{\arraystretch}{1}
\begin{threeparttable}

\textbf{Panel A. Primary outcomes}
\vspace{2pt}

    \begin{tabular*}{\columnwidth}{@{\extracolsep{\fill}}
    >{\columncolor[gray]{0.93}[2pt][2pt]}l
    c
    >{\columncolor[gray]{0.93}[2pt][2pt]}c
    c
    >{\columncolor[gray]{0.93}[2pt][2pt]}c
    @{}}
\toprule
Outcome & Juicy effect & \(p\) & EMM\(^-\) & EMM\(^+\) \\
\midrule
UES-SF total & \(F(1,196)=16.546\) & \cellcolor{blue!20}\(6.871e-5\) & 3.224 & 3.647 \\ 
BeauVis total & \(F(1,196)=24.224\) & \cellcolor{blue!20}\(1.809e-6\) & 4.284 & 5.155 \\ 
Task total & \(F(1,207)=5.925\) & \cellcolor{blue!20}\(1.577e-2\) & 2.703 & 2.832 \\
\bottomrule
\end{tabular*}

\vspace{6pt}
\centering \textbf{Panel B. UES-SF subscales}
\vspace{2pt}

    \begin{tabular*}{\columnwidth}{@{\extracolsep{\fill}}
    >{\columncolor[gray]{0.93}[2pt][2pt]}l
    c
    >{\columncolor[gray]{0.93}[2pt][2pt]}c
    c
    >{\columncolor[gray]{0.93}[2pt][2pt]}c
    @{}}
\toprule
Subscale & Juicy effect & \(p\) & EMM\(^-\) & EMM\(^+\) \\
\midrule
FA & \(F(1,190)=19.699\) & \cellcolor{blue!20}\(1.534e-5\) & 3.005 & 3.490 \\ 
PU & \(F(1,198)=4.163\) & \cellcolor{blue!20}\(4.263e-2\) & 3.433 & 3.714 \\ 
AE & \(F(1,199)=15.228\) & \cellcolor{blue!20}\(1.303e-4\) & 3.139 & 3.616 \\ 
RW & \(F(1,194)=14.883\) & \cellcolor{blue!20}\(1.556e-4\) & 3.332 & 3.758 \\ 
\bottomrule
\end{tabular*}


\end{threeparttable}
\vspace{-16pt}
\end{table}

From qualitative RTA interpretation of participants' preferences, most participants preferred the juicy variants, but not unanimously: 32 (68\%) preferred juicy versions, 12 (25\%) preferred non-juicy versions, and 3 (6\%) reported no difference. 
Among participants who preferred juicy feedback, the most common interpretive and emerging themes were that juicy feedback helps on \textsc{Usability and Readability}\footnote{Note that, as with most qualitative work, we use \textsc{small capitals} to denote themes that we found.} ($n=15$), \textsc{Staying Focus} ($n=4$), \textsc{Fun or Enjoyable} ($n=7$), \textsc{Felt Rewarding} ($n=11$), \textsc{Appealing} ($n=5$), and \textsc{Engaging} ($n=4$). 
In contrast, among participants who preferred non-juicy versions, the emerging themes were that having no or less juicy feedback helped with \textsc{Usability} ($n=7$), \textsc{Readability} ($n=9$), and \textsc{Staying Focus} ($n=4$). 
Here, the RTA revealed a heterogeneous, sometimes double-edged role of juiciness. 
The same themes of \textsc{Staying Focus}, \textsc{Usability}, and \textsc{Readability} appeared in accounts from both participants who preferred the juicy versions and participants who preferred the non-juicy versions.
For the Staying Focus, juicy-preferred participants, for example, explained: ``I feel they [juicy feedback] help make a infographic easier to navigate and keeps you locked in'' (\textbf{P2}), and ``It [juicy feedback] helped wake me up'' (\textbf{P23}); 
While non-juicy-preferred participants wrote: ``I think juicy feedback can often be quite distracting'' (\textbf{P43}) and ``The pings were an annoying distraction'' (\textbf{P40}). 
Related concerns also extended to readability and usability. 
For example, juicy-preferred participants noted: ``The juice feedback to the charts' layout made it easy to understand patterns and comparisons quickly'' (\textbf{P5}), and ``I like the feedback as I know right then if I clicked something and exactly where I clicked it'' (\textbf{P39}); 
While non-juicy-preferred participants commented: ``[Juicy feedback] made it even more difficult to decode the graphs'' (\textbf{P13}) and ``I feel like juicy feedback and embellishments just make navigating the infographic more cumbersome'' (\textbf{P24}). 
This qualitative heterogeneity helps explain the quantitative pattern above: juiciness implications for perceived usability were more individual-dependent, as it clarified action and supported interpretation for some but added cognitive overhead to others. 

\subsection{Evaluating Juiciness in Feedback Types (Study 2)}

Across retained Study 2 trials, mean scores were $M=5.315$ for BeauVis total, $M=3.611$ for UES-SF total, and $M=2.40$ for task performance. 
Raw means by timing coverage suggested that both engagement and aesthetic experience tended to increase as more juicy timing phases were present, whereas task performance showed no comparable monotonic trend. 
Our primary mixed-model analyses (\cref{tab:study2_results}) supported this pattern. 
In contrast, timing coverage did not significantly predict task performance.
These indicate that broader coverage of juicy feedback timing phases (i.e., juicy feedback types) improved participants' subjective experience, but did not yield measurable gains in task accuracy.

To determine which timing placements contributed most to these effects, we fit secondary additive models with binary indicators for \textit{pre-}, \textit{in-}, and \textit{post-}interaction feedback. 
For UES-SF total, only post-feedback showed a significant positive effect, 
with higher estimated marginal means when post-feedback was present than absent. 
Neither pre-feedback 
nor in-feedback 
reached significance. 
Similar in BeauVis total: post-feedback significantly increased BeauVis ratings,
with estimated marginal means of $M=5.515$ when present and $M=5.115$ when absent, whereas pre-feedback 
and in-feedback
were not significant. 
For task performance, none of the three timing indicators was significant. 
This suggests that increasing the temporal coverage of juicy feedback improves perceived engagement and aesthetic pleasure, but that this benefit is driven primarily by post-interaction feedback rather than by pre-interaction or in-interaction cues.

\begin{table}[t]
\caption{Study 2 mixed-model results for timing coverage effects. Panel A reports linear mixed-effects models with timing coverage (\(0\)--\(3\) timing phases) as the fixed effect and participant as a random intercept. Panel B reports corrected additive mixed-effects models with pre-, in-, and post-interaction timing indicators entered simultaneously as fixed effects. 
}
\vspace{-4pt}
\label{tab:study2_results}
\centering
\scriptsize
\setlength{\tabcolsep}{4pt}
\renewcommand{\arraystretch}{1}
\begin{threeparttable}

\textbf{Panel A. Primary models (timing coverage)}
\vspace{2pt}

\begin{tabular*}{\columnwidth}{@{\extracolsep{\fill}}
>{\columncolor[gray]{0.93}[2pt][2pt]}l
c
>{\columncolor[gray]{0.93}[2pt][2pt]}c
c
@{}}
\toprule
Outcome & Test & \(p\) & Slope \(b\) \\
\midrule
UES-SF total & \(F(1,182)=10.772\) & \cellcolor{blue!20}\(1.302e-3\) & 0.110 \\ 
BeauVis total & \(F(1,182)=16.338\) & \cellcolor{blue!20}\(8.652e-5\) & 0.239 \\ 
Task total & \(F(1,182)=1.703\) & \(1.940e-1\) & -0.052 \\ 
\bottomrule
\end{tabular*}

\vspace{6pt}
\centering \textbf{Panel B. Corrected additive timing models}
\vspace{2pt}

\begin{tabular*}{\columnwidth}{@{\extracolsep{\fill}}
>{\columncolor[gray]{0.93}[2pt][2pt]}l
l
>{\columncolor[gray]{0.93}[2pt][2pt]}c
c
>{\columncolor[gray]{0.93}[2pt][2pt]}c
c
@{}}
\toprule
Outcome & Type & Test & \(p\) & EMM\(^-\) & EMM\(^+\) \\
\midrule
UES-SF total & Pre & \(F(1,146)=0.001\) & \(9.701e-1\) & 3.610 & 3.607 \\ 
UES-SF total & In & \(F(1,138)=2.756\) & \(9.918e-2\) & 3.550 & 3.667 \\ 
UES-SF total & Post & \(F(1,139)=8.376\) & \cellcolor{blue!20}\(4.411e-3\) & 3.504 & 3.713 \\ 
\hdashline
BeauVis total & Pre & \(F(1,148)=1.958\) & \(1.638e-1\) & 5.219 & 5.412 \\ 
BeauVis total & In & \(F(1,139)=0.993\) & \(3.209e-1\) & 5.253 & 5.378 \\ 
BeauVis total & Post & \(F(1,140)=9.799\) & \cellcolor{blue!20}\(2.124e-3\) & 5.115 & 5.515 \\ 
\hdashline
Task total & Pre & \(F(1,151)=3.271\) & \(7.252e-2\) & 2.482 & 2.316 \\ 
Task total & In & \(F(1,140)=1.037\) & \(3.102e-1\) & 2.357 & 2.441 \\ 
Task total & Post & \(F(1,142)=0.951\) & \(3.311e-1\) & 2.440 & 2.358 \\ 
\bottomrule
\end{tabular*}


\end{threeparttable}
\vspace{-16pt}
\end{table}


RTA of the juicy feedback type preference responses added details to the quantitative pattern. Participants most often preferred in-interaction feedback ($n=13$), followed by post-interaction feedback ($n=12$) and the full combination of feedback ($n=10$) across all timing phases, whereas fewer participants preferred pre-interaction feedback ($n=3$), the non-juicy baseline ($n=5$), or reported no clear difference ($n=3$). 
Accounts from participants who preferred post-interaction feedback were most consistently organized around the \textsc{rewarding} theme. 
In these responses, it was often experienced as acknowledging user input and making the interaction feel complete. 
The quantitative models suggest that post-interaction feedback contributed most, while the qualitative accounts indicate why: post-feedback was especially likely to be interpreted as confirmation and reward. 
In addition, accounts from participants who preferred the non-juicy baseline were highly consistent with Study 1 in emphasizing the \textsc{staying focus} theme, in which they described added juicy feedback as ``\textit{unnecessary and distracting}'' (\textbf{P2}), ``\textit{annoying}'' (\textbf{P23}, \textbf{P29}), or ``\textit{overwhelming}'' (\textbf{P5}).

\subsection{Evaluating Juiciness with Different Intensity (Study 3)}

In Study 3, results (\cref{tab:study3_results}) provided no reliable evidence that increasing juicy-feedback intensity changed engagement, aesthetic experience, or task performance. 
In the ordered-intensity mixed-effects models, intensity was not a significant predictor of UES-SF total, BeauVis total, or task performance. Secondary categorical mixed-effects follow-up models showed the same null pattern: intensity category did not significantly predict UES-SF total, BeauVis total, or task performance, and no Bonferroni-adjusted pairwise comparison was significant. 
Although the estimated marginal means suggested modest descriptive variation across intensity levels, particularly somewhat higher UES-SF and BeauVis means at the medium condition, these differences were not statistically significant. 

\begin{table}[t]
\caption{Study 3 mixed-model results for ordered juicy-intensity and categorical follow-up analyses. Primary models treated juicy intensity as an ordered numeric predictor (\(0=\) none, \(1=\) low, \(2=\) medium, \(3=\) high) with participant as a random intercept. Secondary follow-up models treated intensity as a four-level categorical factor.}
\label{tab:study3_results}
\centering
\scriptsize
\setlength{\tabcolsep}{4pt}
\renewcommand{\arraystretch}{0.9}
\begin{threeparttable}

\textbf{Panel A. Primary ordered-intensity models}
\vspace{2pt}

\begin{tabular*}{\columnwidth}{@{\extracolsep{\fill}}
>{\columncolor[gray]{0.93}[2pt][2pt]}l
c
>{\columncolor[gray]{0.93}[2pt][2pt]}c
c
@{}}
\toprule
Outcome & Ordered slope \(b\) [95\% CI] & Test & \(p\) \\
\midrule
UES-SF total & \(0.010\ [-0.054,\,0.075]\) & \(F(1,51)=0.105\) & \(7.470e-1\) \\ 
BeauVis total & \(0.005\ [-0.122,\,0.132]\) & \(F(1,51)=0.007\) & \(9.342e-1\) \\ 
Task total & \(-0.097\ [-0.205,\,0.010]\) & \(F(1,52)=3.297\) & \(7.514e-2\) \\ 
\bottomrule
\end{tabular*}

\vspace{6pt}
\centering \textbf{Panel B. Secondary categorical follow-up models}
\vspace{2pt}

\begin{tabular*}{\columnwidth}{@{\extracolsep{\fill}}
>{\columncolor[gray]{0.93}[2pt][2pt]}l
c
>{\columncolor[gray]{0.93}[2pt][2pt]}c
c
@{}}
\toprule
 &  &  & EMMs \\
Outcome & Test & \(p\) & (None/Low/Medium/High) \\
\midrule
UES-SF total & \(F(3,49)=0.357\) & \(7.843e-1\) & 3.376 / 3.308 / 3.415 / 3.376 \\ 
BeauVis total & \(F(3,49)=0.368\) & \(7.767e-1\) & 4.518 / 4.683 / 4.698 / 4.557 \\ 
Task total & \(F(3,50)=2.621\) & \(6.077e-2\) & 2.396 / 2.625 / 2.412 / 2.180 \\ 
\bottomrule
\end{tabular*}


\end{threeparttable}
\vspace{-16pt}
\end{table}

The RTA result was broadly consistent with the null quantitative findings. 
While the medium-intensity condition was most often preferred ($n=10$), low intensity was the next most frequently preferred option ($n=6$), and participants who preferred none ($n=2$) or low intensity often raised concerns about distraction (i.e., the \textsc{Staying Focus} theme). 
Rather than indicating a clear monotonic preference for stronger feedback, these accounts suggest a trade-off between reinforcement and distraction, helping explain why no reliable overall intensity effect emerged in the quantitative models.

\section{Discussion}
\label{sec:discussion}

Our findings let us ask not just whether juiciness transfers from games to data visualization, but where in an interaction it pays off and why. Three study results anchor this discussion: 
First, broadening the temporal coverage of juicy feedback raised engagement and aesthetic experiences, and that gain came almost entirely from post-interaction feedback (Study 2). 
Second, scaling intensity from none to high produced no reliable change in any outcome (Study 3). 
Third, the same qualities, staying focused, usability, and readability, were invoked both by participants who preferred the juicy versions and by those who did not (Study 1). 
Together, these recast juiciness from ``\textit{more-is-better}'' embellishment into a placement- and context-sensitive design material. 
We develop this through the game/work boundary (\cref{sec:discussion1}), content and context (\cref{sec:discussion2}), and a set of design considerations (\cref{sec:discussion3}), before stating limitations (\cref{sec:discussion4}).

\subsection{The Boundary Between ``\textit{Game}'' \& ``\textit{Work}''}
\label{sec:discussion1}

When juiciness \textit{moves} from games to data visualization, what transfers and what shifts? 
Game design centers on engaging player experience through mechanics, visuals, and systems \cite{schell2014art}, whereas interactive visualization is often framed as helping people to make sense of data \cite{JANVRIN201431,robb2018does}.
That sensemaking stance surfaced in our preference data: in Studies 1 and 3, participants judged conditions largely by which one helped them stay focused and extract information, and some ($n=21$, across three studies) who preferred the non- or less-juicy versions often described the effects as interference when they wanted to ``\textbf{get the work done}.'' 
This 
echoes prior literature \cite{singhal2021juicy} describing juicy feedback as ``\textit{polarizing},'' and, when excessive, as a burden. 

Even so, our results suggest that experience design still matters in this work-like setting when it lands at the right moment.
In Study 2, the engagement and aesthetic gains from broader feedback coverage came from post-interaction feedback rather than from pre- or in-interaction cues.
We read this through Norman's action cycle \cite{norman1986user} and Hicks et al.'s notion of supplementary feedback \cite{hicks2018good}: post-interaction juice closes the evaluation phase, confirming the consequence of a completed action.
It also fits evidence that juicy feedback can foster intrinsic motivation; in a non-game VR setting, juiciness on its own satisfied the competence, autonomy, and relatedness needs that \textit{Self-Determination Theory} ties to intrinsic motivation \cite{hicks2019understanding}.
Feedback placed earlier carries the opposite risk; some participants ($n=7$, Study 2) found pre-interaction cues annoying when they fired without an intended interaction, a false-positive signal that undercuts rather than rewards. 
The valence of juiciness thus appears to hinge on where it sits in the interaction loop and on whether a user treats the activity as work or as play. 
The experiential ``\textit{thickness}'' we set out to characterize seems welcome when it confirms an action and intrusive when it pre-empts one.

\subsection{Content \& Context Considerations}
\label{sec:discussion2}

Does every dataset and interaction benefit from juiciness, or do some demand restraint? 
Our accounts suggest its value is contingent rather than uniform, shaped by two factors. 
The first is how much the visualization demands of attention and interaction. 
Some participants ($n=15$, across studies) said juicy elements helped them stay locked in, pointing to value where sustaining attention is hard; others ($n=22$) said the cues helped confirm actions and signal where they were, pointing to value in complex, multi-step interactions. 
Both describe settings that must coax users into an active exploratory role. 
This is a long-standing concern in data-driven storytelling, where news media increasingly use interactive visualization to attract and retain reader attention \cite{lopezosa2022data}, and where the balance between author- and reader-driven narratives is central \cite{lopezosa2022data,segel2010narrative}; it is also the aim of work that frames visualization through empowerment and agency \cite{willett2021perception}.
Reward-oriented juice offers one lever on this balance, nudging readers along an intended path without taking away their control. 
Such contexts already recur in practice, from gameful and feedback-rich designs in visual analytics workflows \cite{sevastjanova2019shall} to student-facing learning-analytics dashboards \cite{bayrak2021powerful}.


The second factor is the seriousness of the content. Some participants ($n=3$, across three studies) preferred an infographic partly because its topic, obesity by country, was ``\textit{interesting}'' (P13), which prompted us to ask whether certain subjects should stay sober and avoid juicy treatment.
We did not manipulate content seriousness and so cannot settle this, but Campos's ethnography \cite{campos2025juiciness} reports the same tension, noting that subject matter conditions whether juicy elements are appropriate and cautioning against overwhelming the core message; work on gamifying a sensitive health simulation likewise finds that playful layers can sit awkwardly with reflective, weighty tasks \cite{tang2023playful}.
Here, a workable reading is that juiciness is most promising where attention and interaction complexity are high and the subject is light, and should be applied more sparingly as content grows heavier.

\subsection{Design Considerations for Juicy Interaction}
\label{sec:discussion3}

From building JuicyVIS and running the studies, we distilled design considerations. 
We offer them as empirically- and literature-informed considerations rather than validated guidelines: each draws on prior work and our findings (\cref{sec:discussion4}).

\textbf{Ensuring granularity in intensity, type, and modality}: 
Prior work in games treats juicy intensity as a design variable and warns that both too little and too much can hurt \cite{kao2024does,singhal2021juicy}. 
Study 3 adds that, within the range we tested, raising juicy intensity from none to high in visualization produced no reliable benefit, while accounts described a trade-off between reinforcement and distraction. 
A moderate default is therefore a sensible starting point, and intensity, type, and modality should stay adjustable so users can tune feedback to their preferences. 
Exposing modality also serves accessibility, echoing multimodal-feedback practice \cite{hicks2018good} and sensory-substitution work in visualization \cite{chundury2021towards}.

\textbf{Match feedback to the interaction it serves}:
A juicy family that suits one interaction can misfit another, so we chose effects to fit each interaction's affordances, for instance a commit-and-confirm pattern for sorting and matched-range cues for filtering (\cref{tab:study1design}). 
Our Study 1 reinforces this: the same feedback was experienced differently across the seven interaction types \cite{yi2007toward}, so cues should follow what an interaction means rather than a single juicy template applied everywhere \cite{campos2025juiciness}.

\textbf{Calibrating juiciness to content and context}:
As \cref{sec:discussion2} argued that, the same feedback does not fit every setting equally. 
It is most valuable where attention or interaction complexity is high, and should be restrained for serious or weighty subjects, where playful layers can undercut the message \cite{campos2025juiciness,tang2023playful}. 
Treating content seriousness and task demand as first-class design inputs helps decide how much juice a visualization should carry.

\textbf{Providing confirmation of successful interaction}:
Confirming a completed action is a familiar goal, tied to Norman's evaluation phase \cite{norman1986user} and to Elmqvist et al.'s call for immediate feedback that rewards interaction \cite{elmqvist2011fluid}. 
Our contribution is to pin down where it pays off: in Study 2, post-interaction feedback drove the engagement and aesthetic gains, whereas pre- and in-interaction cues did not, and pre-interaction cues that fired without a committed action read as annoying false positives. 
Designers should prioritize feedback that closes the loop after an action and avoid exceeding cues that anticipate one.

\textbf{Reinforcing designer-intended interaction}:
Juicy feedback can reward particular actions and steer attention, letting designers express intent through when and where feedback is enabled. 
Set against the balance of author- and reader-driven narrative \cite{lopezosa2022data,segel2010narrative} and visualization framed through empowerment \cite{willett2021perception}, the aim is to guide users along an intended analytic path while leaving them in control.

\textbf{Avoiding disruption to the core functionality}:
Juicy should not crowd out the work of reading data. 
Embellishment can aid memorability yet carries comprehension trade-offs \cite{bateman2010useful}, and Campos's ethnography flags distraction and overstimulation as genuine risks \cite{campos2025juiciness}.
Study 1 sharpens the point: the same features helped some participants and hindered others, so disruption is partly individual.
Visually, cues should not obscure marks or controls; technically, they should not add latency; and when either occurs, the feedback should be reducible or removable, which loops back to adjustability. 

\textbf{Preserving and sharing juicy interactive data visualization}:
Because juicy feedback is fine-grained and unfolds in real time, it resists static description. 
Recording and sharing the underlying designs and parameters supports transparency and reuse.
Our Juice Elements Palette (\cref{tab:juicepalette}) is a step toward a reproducible, inspectable vocabulary for a space documented so far mostly by example.

\subsection{Limitations \& Future Work}
\label{sec:discussion4}

We designed and refined the JuicyVIS prototypes and the considerations between ourselves, informed by theory, pilots, and our studies, but without validation by external design professionals or domain experts; we therefore present them as considerations to be tested, and see structured experts critique and iterative refinement as the natural next step. 

Other limits temper our claims. 
Our online Prolific samples skew Western and English-speaking. 
Each study was a single session, leaving novelty decay untested. 
Our stimuli and intensity presets are one principled instantiation, not an optimum, and we did not vary content seriousness, so \cref{sec:discussion2} stays a hypothesis. 
We also studied the three dimensions in isolation rather than their interactions. 
Each of these marks an opening for future work.

\section{Conclusion}

Interactive visualizations often ask users to act yet say little back. 
We introduced \textit{JuicyVIS}, a theory-informed operationalization of juicy feedback that translates game studies ideas into a visualization design space grounded in interaction type, feedback timing, and feedback intensity. 
Across 26 controlled prototypes and three exploratory mixed-methods online studies, 
juicy feedback generally improved engagement and aesthetic experience, showed a smaller task advantage 
and the strongest gains when feedback followed and an interaction as a reinforcing cue; stronger intensity alone did not reliably help, and accounts repeatedly revealed a tension between reinforcement and distraction.
These results position juicy feedback as a design material for how interaction ``\textit{\textbf{feels}},'' more than mere embellishment, to be applied deliberately and with attention to timing, task, and interpretive load.
We hope JuicyVIS offers a vocabulary, a set of exemplars, and an empirical starting point for visualization feedback that is not only functional, but experientially legible and alive.

\acknowledgments{%
This work was supported in part by the National Science Foundation \#2213757.
During the development of JuicyVIS prototype (described in \cref{sec:design}), \textit{OpenAI Codex} (in \textit{Visual Studio Code}) was used only for debugging; 
No artificial intelligence (AI), large language models (LLMs), or similar systems were used to generate the design, prototyping, research, analysis, text, figures, or references in this work.
}

\section*{Supplemental Materials}
\label{sec:supplemental_materials}

The interactive demo of JuicyVIS prototypes can be found at \href{https://juicy-vis-10aac.web.app/}{juicy-vis-10aac.web.app/\(\nearrow\)}, and its original codes can be found at \href{https://github.com/revisit-games/juicyvis}{github.com/revisit-games/juicyvis\(\nearrow\)}.
The codes of deployed JuicyVIS online studies can be found at \href{https://github.com/revisit-games/JuicyVIS-studies}{github.com/revisit-games/JuicyVIS-studies\(\nearrow\)}.
The de-identified trial-level datasets for all three studies, SPSS analysis syntax used to reproduce the reported models, and corresponding output files can be found at \href{https://osf.io/bs4cz/}{https://osf.io/bs4cz/\(\nearrow\)}.

\bibliographystyle{abbrv-doi-hyperref}

\bibliography{REFS}








\end{document}